\documentclass[twocolumn,showpacs,preprintnumbers,amsmath,amssymb,prl]{revtex4}
\usepackage{graphicx} % Include figure files
\usepackage{subfigure}
\usepackage{dcolumn} % Align table columns on decimal point
\usepackage{bm} % bold math
\usepackage{color}
\usepackage{ulem}
\usepackage{subfigure}
\usepackage{float}

\begin{document}

\preprint{APS/123-QED}
\title{On the relation between linear stability analysis and mean flow properties in wakes.}

\author{Benjamin Thiria}
% \email{thiria@ensta.fr}
\affiliation{Unit\'e de M\'ecanique, Ecole Nationale Sup\'erieure
de Techniques Avanc\'ees, Chemin de la Huni\`ere, 91761 Palaiseau
Cedex, France.}%

\author{Gilles Bouchet}

\affiliation{Institut de M\'ecanique des Fluides et des Solides, 2 rue Boussingault, F67000 Strabourg, France.}%

\author{Jos\'e Eduardo Wesfreid}

\affiliation{Physique et M\'ecanique des Milieux H\'et\'erog\`enes
(UMR 7636 CNRS-ESPCI, P6, P7), Ecole Sup\'erieure de Physique et
Chimie Industrielles de Paris , 10 rue Vauquelin, 75231 Paris Cedex
5,
France.}%

\date{\today}% It is always \today, today,

\begin{abstract}

In recent studies on wake stability \cite{Thiria07,Barkley06,Pier02}, it has been observed that a simple linear stability
analysis applied to the mean flow instead of the basic flow, could
give an accurate prediction of the global mode selected frequency,
although these phenomena are strongly non-linear. In this letter, we
study the transient regime between the stationary (so called basic
state) and unstationary solutions of the wake of a circular cylinder
at Re=150. We show that the shift of the global frequency as a
function of time due to strong non-linear effects, can be
interpreted by a continuous mean flow correction induced by the
growth of the instability. We show that during this transient
regime, the mean state as a function of time plays the role of an
instantaneous basic state on which the global frequency can be
determined linearly.

\end{abstract}

\maketitle

\begin{figure}
\begin{center}
\includegraphics[width=0.50\textwidth]{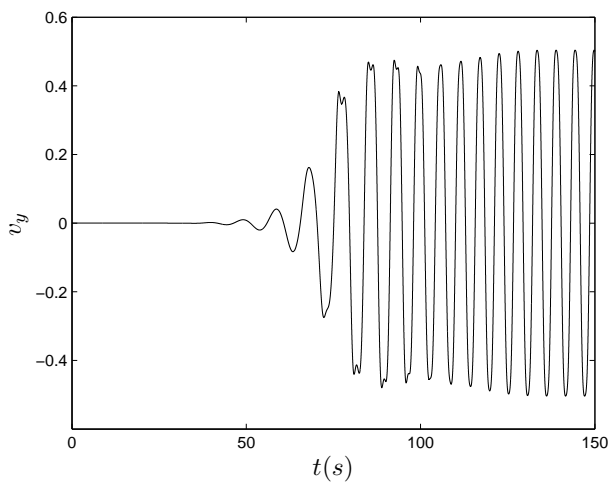}
\end{center}
\caption{Time evolution of the transversal velocity component
measured at $y=0$ during the transition from the steady to saturated
state.} \label{Transit}
\end{figure}

\begin{figure}
\begin{center}
\includegraphics[width=0.50\textwidth]{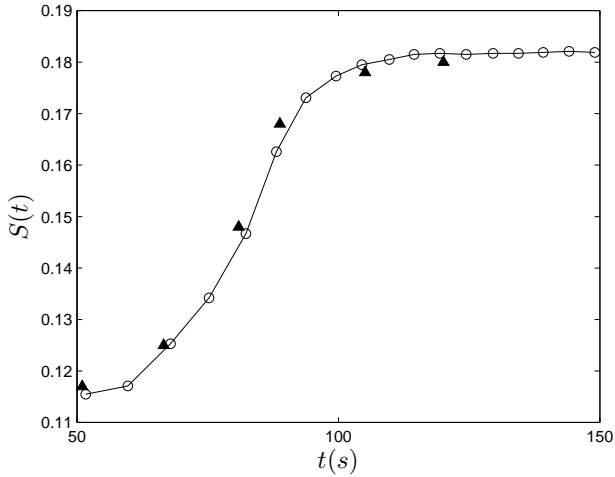}
\end{center}
\caption{Instantaneous Strouhal number as a function of time during
the transient regime and comparison between numerical and
theoretical results (which will be discussed later). Values denoted
by $\blacktriangle$ correspond to global frequencies obtained by
applying the linear saddle point criterion \cite{Chomaz91} on the
instantaneous mean flow.} \label{freqt2}
\end{figure}

It is known that most unstable flows such as shear layers, jets and
wakes present a strong non-linear dynamic even near the threshold of
instability. For synchronized open flows, like bluff body wakes,
these behaviors can be described by a Landau or Ginzburg-Landau
model \cite{Chomaz05}, directly linking the non-linear frequency and
amplitude of the perturbation with the distance between the control
parameter and its critical value at the threshold
\cite{Provansal87}. However, recent results have shown that the best
theoretical predictions describing the behavior of such flows were
based on a fully linear theory but applied to a mean state (i.e.
time averaged unstationnary flow) instead of the usual basic steady
state
 \cite{Pier02}. Equivalent results using this technique had already
been obtain by \cite{Triantafyllou86} on directly measured mean
velocity profiles but no comment was made of the fact that the mean
flow gave good frequency predictions. This important result have
also been confirmed recently in \cite{Barkley06} using this time a
global linear stability analysis. With this analysis, it was shown
that the solution obtained from the mean flow was marginally stable
(i.e the growth rates of the eigenvalue computed were found to be
almost zero). Similar conclusions have been obtained for forced open
flows (wake behind an rotating oscillating cylinder)
\cite{Thiria07}. In this case, a significant change in the mean flow
induced by the forcing conditions, have been observed. The
consequences of this mean flow distortion were that the stability
properties of the forced flow, and so the selected spatial modes,
changed, following scaling laws as a function of the forcing
parameters as can be observed for natural wakes \cite{Zielinska95}
with the Reynolds number. In recent studies, Thiria \& Wesfreid
\cite{Thiria07} have conjectured that the strong non-linearities
generated by these unstable flows are concentrated in this mean flow
distortion. The mean state would play the role of a new basic state
where linear modes (defined by their frequency, amplitude and
spatial shape) are selected as a function of its properties. More
precisely, it would mean that any changes in the mean configuration
of the flow would be equivalent to a change in the selected mode,
which is selected linearly and fixed by the mean state. In the
present letter, we study the transient regime of a flow past a
circular cylinder at $Re=150$ from its steady to saturated state:
the flow is first forced to be symmetrical about the axis $y=0$, a
perturbation is then applied and the flow evolves and converges in
time to its unstationary solution. We show that the non-linear shift
of the frequency during this transient regime
\cite{Provansal87,Thompson04}, can be explained by a continuous
distortion of the mean flow as a function of time, generated by the
growth of the perturbation, and
that this frequency can be predicted instantaneously by a linear stability analysis.\\
The flow past the circular cylinder at $Re=150$ and then the dynamic
of the transient regime have been obtained by numerical computation
using the spectral finite element code $NEKTON$ \cite{patera84}
simulating an unconfined $2D$ wake. The cylinder has a diameter of
$d=1$ and the upstream velocity is $U_{0}=1$, so the viscosity $\nu$
is set to get a Reynolds number of $Re=150$. The figure
\ref{Transit} displays a typical time evolution of the transversal
fluctuating velocity component $v_{y}$ during this transition
showing that the selected period decreases to lower value as time
increases (i.e. shift to higher frequency as a function of time).
The instantaneous Strouhal number $St(t)=f(t)d/U_{0}$ has been
extracted using a wavelet transform applied to the velocity signal
$v_{y}(t)$ displayed in figure \ref{Transit}. The evolution of
$S(t)$ is displayed in figure \ref{freqt2}.

\begin{figure}
\begin{center}
\includegraphics{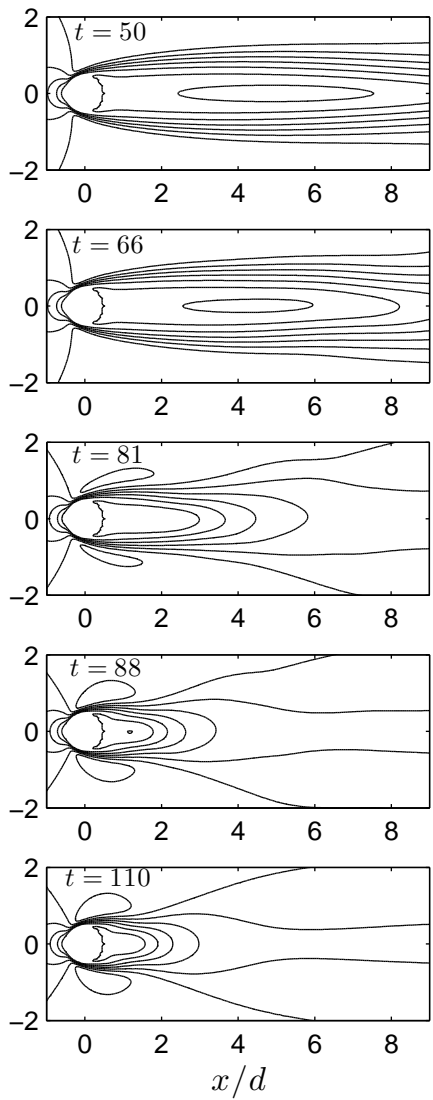}
\end{center}
\caption{Evolution of the mean flow as a function of time during the
transitory represented by lines of isovalues of the velocity
modulus. The first case corresponds to the basic flow while the last
case corresponds to the time averaged flow of the fully developed
street at $Re=150$.} \label{meanflowt}
\end{figure}

As can be seen, the Strouhal number evolves from $St=0.116$ (the
first frequency selected by the wake) to $0.182$. This typical
evolution illustrates how the global selected frequency is shifted
due to the strong non-linear effects as the growth of the
instability increases. \\
We also studied the evolution of the time dependant constructed mean
flow by an ensemble average as a function of time during this
transient regime . The information concerning this local mean state
were obtained by computing $100$ transient regimes for the entire
flow field $(u(x,y),v(x,y))$ triggered each time by a controlled
initial perturbation with a different phase. These different
realizations then allowed access, by an ensemble average, to the
instantaneous mean flow as a function of time \footnote{A linear
eigenfunction is used to perturb the base flow. This eigenfunction
is computed by subtracting the base flow (obtained by forcing the
symmetry around the x-axis of the cavity at y=0) to the full flow
pattern in the linear regime (starting from the basic flow, the
numerical noise is sufficient to start the instability). This
procedure is repeated at $t + n  t_{i}$ for $t_{i} =n   / 100$
(being the period in the linear regime) and we finally obtain $100$
eigenmodes, each with exponentially growing amplitude and different
phase. We normalize the amplitude of the eigenmodes to $10^{-3}$ and
thus have a wide base of initial perturbations.}. This is displayed
in figure \ref{meanflowt}. The evolution of the flow during the
transient regime from the initial symmetric basic steady state to
saturated state shows a drastic change in its mean properties due to
strong non-linear effects as already observed \cite{Zielinska97} and
shows clearly that the mean spatial shear layers are very different
between the beginning and the end of the transient regime where the
size of the recirculation bubble is reduced. This first observation
supposes strongly that there are no obvious reasons to affirm that
the instability which is observed for the developed wake, can be
explained by simply studying the unstable modes on the basic state.

\begin{figure}
\begin{center}
\includegraphics[width=0.50\textwidth]{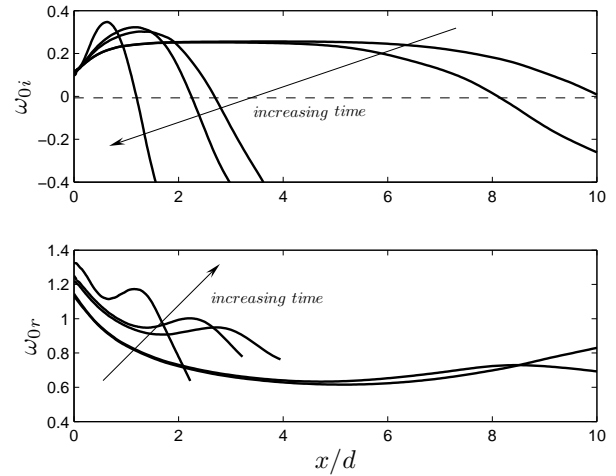}
\end{center}
\caption{Spatial evolution the local absolute growth rate
($\omega_{0i}$) and local pulsation ($\omega_{0r}$) as a function of
the longitudinal coordinate $x$ as time increases during the
transitory (t=50, 66, 81, 88 and 110).} \label{omegaietrx}
\end{figure}

\begin{figure}
\begin{center}
\includegraphics[width=0.50\textwidth]{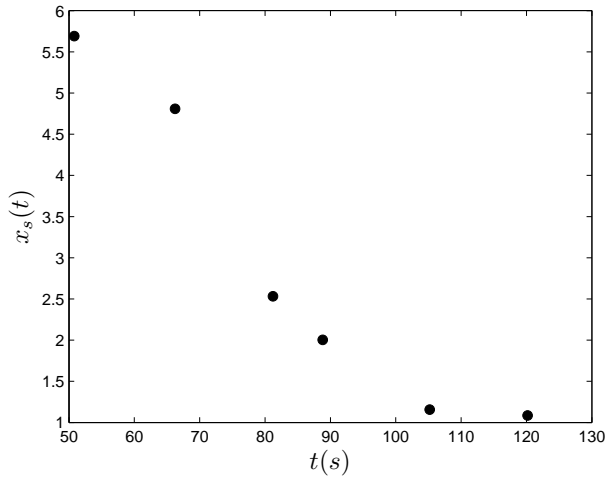}
\end{center}
\caption{Evolution of the saddle point location as a function of
time during the transient regime of the wake.} \label{xstime}
\end{figure}

Then, the local stability properties of the mean flow, for each
time, were obtained by solving numerically the inviscid
Orr-Sommerfeld equation for the streamfunction $\Psi(x,y,t)= \int
_{0}^{y} U(\eta) d \eta + \psi(x,y,t)$, where $\psi(x,y,t) =
\mathcal{R}e \lbrace \phi(y) e^{i(kx-wt)} \rbrace$:

\begin{equation}
(k U(y) -\omega)(\phi''-k^{2} \phi) -kU''(y) \phi =0
\end{equation}

with boundaries conditions $\phi(-\infty)=\phi(+\infty)=0$, and
where $U(y)$ is the local velocity profile depending on the
transversal coordinate $y$, $k$ and $\omega$ are respectively the
complex wave number and the complex frequency of the perturbation
and $\psi$ the associated eigenfunction. It has been shown,
\cite{Triantafyllou86,Thiria07} that the use of an inviscid equation
instead of the complete Orr-Sommerfeld equation was sufficient to
determine the local properties of the flow even for low Reynolds
numbers. Since we want to determinate the theoretical predicted
global frequency, one has to differentiate a convective from an
absolute local instability and then look for solutions verifying
$\mathcal{D}(\omega,k)=0$ and $\left (\frac{\partial
\omega}{\partial k}\right)_{k_{0}}=0$ \cite{Huerre98} where
$k_{0}=k_{r}+ik_{i}$ and $\omega_{0}=\omega_{r}+i\omega_{i}$ are
respectively the complex absolute wave number and frequency. One has
to note that, even if the global linear stability analysis as used
in \cite{Barkley06} is the more accurate, the choice of using local
stability analysis in this paper is deliberate. It has been used in
order to physically understand precisely the evolution of the
stability properties as a function of time which is
not possible with the technique cited above.\\
Local properties of the wake (absolute frequency $\omega_{0r}(x)$
and absolute growth rate $\omega_{0i}(x)$) during the transient
regime are displayed in the figure \ref{omegaietrx} for the five
instances in time chosen in figure \ref{meanflowt}. As can be seen,
the absolute region ($\omega_{0i}>0$), initially extends over 10
cylinder diameters when the wake is in its basic state decreases
quickly with time and reaches its final value at saturation. This
strong evolution can also be seen on the spatial distribution of
absolute local frequencies $\omega_{0r}(x)$. As time increases,
variations of $\omega_{0r}$ are faster and their local values
globally increase in
the absolute region. \\
The global frequency was then determinate by applying the saddle
point criterion on the numerical results based on the analytic
continuation of $\omega_{0}(x)$ in the $x$-plane and given by
$\omega_{g}=\mathcal{R}e(\omega_{0}(x_{s}))$ with
$\frac{d\omega_{0}}{dx}(x_{s})=0$ and where $x_{s}$ is a complex
number and $\omega_{g}$ represents the global frequency selected by
the wake deduced from local properties \cite{Chomaz91}. The method
used to calculate this global frequency is detailed in
\cite{Thiria07}. Finally, we compare the numerical results with
those of our theoretical prediction. These are plotted in figure
\ref{freqt2} and we see clearly that a simple linear criterion
closely follows the frequency obtained with our numerical
simulation. Finally, figure (\ref{xstime}) shows the evolution of
the real part of the saddle point $\mathcal{R}e(x_{s})$ giving the
selected frequency as a function of time and indicates that the
critical point responsible for the selection of the global frequency
moves back close to cylinder. One can note that this evolution
follows those observed for the recirculation zone observed in figure
\ref{meanflowt}.
 \\

Thus, in this letter, we show the fundamental role played by the
mean flow correction in the stability properties in fluid mechanics.
The main important comment to be made is that the basic flow is not
the right flow to be considered when predicting stability properties
as it differs strongly from the real observed flow. The B\'enard-von
K\`arm\`an instability is driven by the two unstable shear layers
located on both sides of the body and it is clear that they are
affected by the non-linear mean flow correction. The consequence is
that the real flow has no link with the basic one. It seems that the
mean flow, even if it is the result of a temporal averaging, is the
closest flow to the real one which can be understood if we consider
that modes are selected linearly. The mean flow correction changes
the equilibrium position (or stable fixed point) of the system by
modifying the size and the intensity
of the absolute region. \\

One has to note that theoretical arguments have to be put forward to
establish clearly the validity of using linear stability on time
averaged flow and this problem is now one of the preoccupation of
the community \cite{Thiria07,Barkley06,Pier02}. However, the fact
that stability properties are related to the mean state gives a new
perspective in the study of such flows. For example, it is now
possible to understand global behavior and mode selection in
controlled or forced flows, for which it is established that the
simple fact of step in a flow changes its mean state
\cite{Thiria07}, but also determinate properties of large scales of
turbulent flows, which should be dominated by the mean shear.

%\newpage %Just because of unusual number of tables stacked at end
\bibliographystyle{unsrt}

% Produces the bibliography via BibTeX.

\end{document}